# *Monolithic piezo-magnonic-MEMS for efficient modulation of RF signals*

M. Cocconcelli, A. Angotti, P. Florio, N. Pellizzi, F. Maspero, R. Bertacco

*Dipartimento di Fisica, Politecnico di Milano, Via G. Colombo 81, 20133 Milano (Italy)*

**Abstract:**

Compact, low-power, analog RF components are essential for next generation microwave electronics and wireless systems. We demonstrate an all-electric integrated piezo-magnonic microelectromechanical system that enables efficient voltage control of GHz spin wave signals via magnetoelastic coupling. Exploiting the large strain on a CoFeB magnonic waveguide integrated on a silicon bridge with piezoelectric actuation, very large values of the magnetoelectric field (up to 30 mT at 30 V) are obtained, thus achieving reversible phase and amplitude control of propagating spin waves. In the static regime we achieve either up to $4\pi$ phase modulation or ≈50 dB amplitude attenuation with drive voltages below 20 V at 7 GHz. Leveraging the bridge's first bending resonance (≈17 kHz) yields resonant enhancement of the phase modulation efficiency. This allows to achieve $2\pi$ phase swing with just 2.2 V drive and power consumption ≈6 μW. Our results highlight piezo-magnonic-MEMS as a promising new class of devices for reconfigurable RF front ends and analog signal processors.

Wireless communication is one of the primary technological drivers of modern society. With the rise of 5G—and the upcoming deployment of 6G—networks and the expansion of the Internet of Things (IoT), the volume of transferred and processed data is increasing exponentially.[1] More generally, RF applications are becoming increasingly pervasive across domains such as monitoring, automotive systems, medicine and security.[2,3] Magnonics, the field of magnetism dealing with spin-waves (SW) and related quasi-particles (magnons), has emerged as a promising technology for next-generation microwave systems. In fact, it can offer highly innovative solutions by leveraging the intrinsic tunability of spin waves, based on the strong dependence of the band dispersion on the magnetic bias field $H_e$.[4,5,6,7,8,9] Magnonic devices are ideal for edge processing of data encoded in RF signals since they allow for a direct physical transfer of an electromagnetic signal into SW without any frequency conversion. Furthermore, magnonics is highly versatile, enabling a broad range of computational approaches. Spin-wave platforms have been proposed and demonstrated for Boolean logic,[10] neuromorphic computing,[11] and reservoir computing.[12,13] Crucial for this is the wave-nature and nonlinearity of SW, providing computing functionalities via interference and nonlinear interaction of the magnons modes.[14]

That said, in the near term, practical applications of magnonics are expected in RF signal processing. Recently, a SW band-pass filter for 6G communication embedded in a radio receiver has been demonstrated, exhibiting features that make it appealing for application in RF-front end modules.[15] The work by Devitt et al. on one side demonstrates the high potential of magnonics, but on the other side evidences the obstacles for a wide spread of this technology, mainly related to the integration in standalone and reconfigurable microelectronic devices. Suitable strategies for integrating tunable, low-power sources of the bias magnetic field $H_0$ are still missing, as well as methods for varying said field to tune the RF frequency response. In a recent paper we reported on a first example of standalone integrated magnonic device, featuring integrated permanent micromagnets[16] coupled to magnetic flux concentrators which allow to tune the bias field on a CoFeB waveguide.[17] In that case the tuning was done by design, setting the distance between the permanent micromagnets and the flux concentrators, and it was suggested that the real time reconfigurability could be implemented using a Micro-Electro-Mechanical Systems (MEMS) to vary that distance in a hybrid magnonic-MEMS device.

Here we propose an alternative approach that uses magnetocrystalline anisotropy to define the effective field $H_0$ and leverages magnetoelastic coupling to modulate it, thereby enabling dynamic tunability of the device frequency response.[18] We demonstrate a highly efficient static and dynamic phase/amplitude modulation in compact and low-power devices realized with planar technology suitable for integration in conventional electronic chips. Crucial to this is the fabrication of a magnonic waveguide directly on a suspended piezo-bridge, allowing to produce extremely large values of strain and to exploit the mechanical resonance to amplify the modulation efficiency.

**Concept of SW modulation via magnetoelastic coupling in suspended piezo-bridges**

Inverse magnetostriction—the generation of an effective magnetic field in a ferromagnet under stress—is a long-established route to electrically tune magnetic properties.[19] The standard approach couples a piezoelectric actuator to a ferromagnet so that electrically generated strain produces a magnetoelastic effective field $H_{ME}$.[20,21] For cubic systems, such as CoFeB used for waveguides in the present work, $H_{ME}$ is given by:

$$H_{ME} = -\frac{2}{\mu_0 M_s} \begin{bmatrix} B_1 m_x \varepsilon_{xx} + B_2(m_y \varepsilon_{xy} + m_z \varepsilon_{xz}) \\ B_1 m_y \varepsilon_{yy} + B_2(m_x \varepsilon_{xy} + m_z \varepsilon_{yz}) \\ B_1 m_z \varepsilon_{zz} + B_2(m_x \varepsilon_{xz} + m_y \varepsilon_{yz}) \end{bmatrix} \quad (1)$$

where $\mu_0$ is the vacuum permeability, $M_s$ the saturation magnetization, $B_1$ and $B_2$ the magnetoelastic coefficients, $m_{x,y,z}$ the reduced magnetization components, and $\varepsilon_{ij}$ the strain tensor components. For CoFeB ($B_1, B_2 \sim 10^6 \, J \, m^{-3}$, $M_s \approx 1.27 \times 10^6 \, A \, m^{-1}$) a field change of ~10 mT requires strain on the order of $10^{-3}$. Such values are difficult to achieve in macroscopic substrate-clamped structures.[22] Partially released piezoelectrics can reach this regime, but typically only for narrow structures limiting the cross section of the magnonic conduit and thus the transmitted spin-wave power.[23] Reported magnetoelectric coupling coefficient are quite high (1.3 mT/V) but the applied voltages cannot often exceed a few volts, leading to magnetoelectric fields < 5 mT.

To overcome these limitations, in this work we use MEMS made of suspended piezoelectric bridges with a central constriction acting as a strain concentrator.[24] As shown in **Fig. 1a**, each silicon bridge incorporates four lead zirconate titanate (PZT) patches connected in parallel. A voltage between the top and bottom electrodes sandwiching a 2 μm PZT layer induces vertical bending via the $d_{31}$ coefficient (inset of **Fig. 1c**). At the constriction, applying $V_p = 10V$ to the piezo produces a tensile strain of ≈ 1.5x10$^{-3}$ over an area of ≈50 × 50 μm$^2$, sufficient to accommodate a 20 μm-wide CoFeB magnonic waveguide (40 nm thick) with meander antennas (red and yellow regions in Fig. 1b).

The measured voltage-dependent out-of-plane displacement (Δz) of the constriction is reported in **Fig. 1c**. A linear behavior is found between 5 and 25 V, where the slope is ≈ 0.3 μm/V, while the deviations from linearity at low and high voltages depend on the hysteretic behaviour of the piezo and on the pooling procedure. The spatial distribution of the strain components at $V_p = 10$ V, as resulting from COMSOL Multiphysics simulations with Δz corresponding to the experimental value for 10 V applied, is shown in **Fig. 1f**. Between the antennas a nearly uniform tensile strain $\varepsilon_{xx} \approx 1.5 \times 10^{-3}$ is found, while $\varepsilon_{yy} \approx -4 \times 10^{-4}$ and $\varepsilon_{zz} \approx -1 \times 10^{-4}$. Among shear terms, only $\varepsilon_{xy}$ reaches the $10^{-4}$ range, while others are negligible. To evaluate the magnetoelastic field, the equilibrium magnetic configuration must be defined. The CoFeB film was grown under a 27 mT field to induce uniaxial anisotropy along the transverse (y) direction. Angular resolved ferromagnetic resonance (FMR) measurements on reference films yield a value of the anisotropy constant $K_{an} = 2 \times 10^4$ J m$^{-3}$, corresponding to an anisotropy field $H_{an} = 2 \cdot K_{an}/\mu_0 M_s = 33$ mT. Upon application of a positive external field H$_e$ along y, the transverse anisotropy overcomes shape anisotropy and stabilizes an almost uniform remanent magnetization along $y$, as confirmed by micromagnetic simulations reported in **Fig. 1g**. From Eq. 1, neglecting terms containing m$_x$, m$_z$, $\varepsilon_{xy}$ and $\varepsilon_{yz}$, the magnetoelastic field for small applied voltages can be written as $\boldsymbol{H_{ME}} = -\frac{2B_1(m_y \varepsilon_{yy})}{\mu_0 M_s} \hat{y}$. Noteworthy, as both $B_1$ and $\varepsilon_{yy}$ are negative, H$_{ME}$ opposes the remanent magnetization (see Fig. 1d). Panel 1h shows a color map of the magnetoelastic field obtained from the full Eq. 1, using B$_2$ = - B$_1$ = 1.6·10$^7$ J m$^{-3}$ consistent with data in literature and yielding a good fit to the broadband spectroscopy data.[23,25,26] A nearly uniform negative transverse field $\mu_0 H_{ME,y} \approx -10$ mT at 10 V in the area between the antennas is found (**Fig. 1h**). In the Damon–Eshbach (DE) configuration used in this work (wave vector **k** of SWs perpendicular to the equilibrium magnetization M$_0$), the **H**$_{ME}$ generated by strain therefore reduces the total transverse effective field and shifts the spin-wave dispersion to lower frequencies.

The effect on SW transmission is illustrated in **Fig. 1e**. Shown are the analytical dispersion[27] for DE SW in the region between the antennas, calculated for H$_e$ = 0 using the uniform **H**$_{ME}$ and **M** estimated above, in case of $V_p = 0$ (no strain) and 10 V (tensile strain). The excitation efficiency of the meander antennas is also reported with dashed line. The whole frequency band related to the fundamental mode, defined by the first peak in the antenna excitation efficiency (grey area in panel 1e), shifts downwards by ~0.6 GHz while producing a change in wave vector at fixed excitation frequency. This voltage-controlled wave-vector variation enables the tuning of phase accumulation, $\Delta \phi = \Delta k \cdot d$, as illustrated in Fig. 1e for a central frequency of 7 GHz that will be used for the experimental validation of this concept. By further increasing the piezo voltage, both the negative y-component H$_{MEy}$ and also the x-component $H_{MEx} = -\frac{2B_2(m_y \varepsilon_{xy})}{\mu_0 M_s}$ tends to break the conduit into magnetic domains (see Figure 3) suppressing SW propagation. As detailed in the following, for low voltages the device behaves as a phase shifter, while for higher voltages it implements a RF switch.

**Experimental determination of the magnetoelectric coefficient from SW propagation**

Figure 2 reports the broadband spectroscopy characterization of DE SW in the CoFeB conduit as a function of the applied transverse field $H_e$, for $V_p = 0$ V, measured on a device with 30 μm edge-to-edge distance between the meander antennas. In panel 2a the color map of the imaginary part of the scattering parameter $S_{21}$ is reported as a function of frequency and $H_e$ for $V_p = 0$ V. The map is almost symmetric with respect to $H_e$, apart from the non-reciprocity of DE SW which reflects in the slightly reduced intensity of the left part. Interestingly, the signal persists also for zero applied field, because of the transverse anisotropy set during CoFeB deposition. The coercive field measured by microMOKE is 1.5 mT. This is enough to operate the device without external applied magnetic fields, i.e. in a "standalone" configuration.

In panel 2b we show with continuous line the Im($S_{21}$) spectra recorded at $V_p = 0$ V, upon removal of the direct electromagnetic coupling for some values of the applied magnetic field from 1 to 25 mT. These spectra have been compared with another set recorded at variable $V_p$ and fixed $H_e$=25 mT, to calibrate our system and find a value of the converse magnetoelectric coefficient $\alpha_{ME} = \mu_0 \cdot dH_{ME}/dV$. The dashed curves reported in Figure 2 correspond to the Im($S_{21}$) recorded at $H_e$=25 mT and $V_p$ values (reported within brackets) that allow us to reproduce the continuous lines. In fact, the application of a positive $V_p$ creates a negative y-component of $H_{ME}$ which is equivalent to a reduction of the external field. For each couple of spectra, the magnitude of $H_{ME}$ created by that specific $V_p$ at $H_e = 0$ mT can be obtained from the difference between 25 mT and the $H_e$ value needed to reproduce the same spectrum at $V_p = 0$ V. The estimated $H_{ME}$ increases almost linearly with $V_p$ reaching 24 mT at 24 V. Slight deviations below 10 V reflect the nonlinear displacement–voltage relation (Fig. 1c). The static converse magnetoelectric coefficient is $\alpha_{ME} \sim 1$ mT/V, comparable to the best values reported in the literature.[23,28,29,30] However, the magnetoelectric fields used in this work (up to ~30 mT) represent record values for spintronic and magnonic devices with relatively thick ferromagnetic layers (tens of nm).[31] Larger values have been reported only in case of ultrathin ferromagnets and in the context of voltage controlled magnetic anisotropy.[32,33,34,35] Note that, since the MEMS can tolerate drive voltages up to 100 V, high field strengths on the order of ≈100 mT could be attained.

**Static phase and amplitude modulation of the spin-wave signal**

**Figure 3a** shows a color map of the Im($S_{21}$(f)) measured at a fixed external applied field of 25 mT, analogous to that shown in Figure 2a; here, however, the horizontal axis is the piezo drive $V_p$ (from 0 to 30 V in 2 V steps) rather than the external magnetic field. The dashed curves in Figure 2b correspond to sections of this color map at selected values of $V_p$. Increasing $V_p$ induces a progressive downward frequency shift of the bands associated with the fundamental and first width modes, consistent with the negative y-component of the magnetoelastic effective field. A pronounced frequency shift of ≈3 GHz is observed, demonstrating highly efficient electrical modulation of spin-wave propagation in the device.

**Figure 3b** reports the corresponding map measured at zero external magnetic field, i.e. in the "standalone" configuration used in the following. The downward bending of the isophase lines induced by the magnetoelectric field is clearly visible up to about 12 V. Above this value the magnitude of the SW signal drops, while the isophase lines become almost horizontal. Two regimes can be distinguished: (i) $V_p < 12$ V - the dispersion shifts while the SW amplitude remains essentially unchanged; (ii) $V_p > 12$ V - the SW amplitude rapidly diminishes, preventing accurate phase determination. This clearly emerges from **Figure 3c**, were we show $Im(S_{21}(f))$ spectra for $V_p$ between 0 and 20 V. A substantial downward frequency shift (~1.5 GHz) of the SW band is seen, looking at the position of the FMR frequency which corresponds to the low-frequency onset of the oscillations.

**Figure 3d** shows the corresponding unwrapped phase $\Phi_{SW}$. Within the spin-wave band $\Phi_{SW}$ decreases nearly linearly with frequency (apart from small residual undulations due to interference with the direct electromagnetic signal), as expected from $\Phi_{SW}(f) = k(f) \cdot d$, where $k(f)$ is the dispersion relation inverse and d the antenna separation, considering that $k(f)$ is almost linear in this frequency range. Noteworthy, we observe that the SW phase at fixed frequency is largely modulated by $V_p$. Figure 3e shows the phase variation $\Delta\Phi$ at 7 GHz - a quite central frequency for the SW bands in the 0-14 V range - induced by the application of an increasing $V_p$. The phase reaches a maximum shift of ≈750° and exhibits an approximately linear dependence above ~4 V (near the coercive voltage of the piezo) corresponding to a phase modulation coefficient $\gamma = \Delta\phi/\Delta V \approx 62$ deg/V.

The overall behaviour is summarized in **Figure 3f**, reporting the amplitude and phase of $S_{21}$ at 7 GHz for different $V_p$. In the 0-12 V range the device produces a highly efficient phase modulation at essentially constant amplitude, as expected for an ideal phase shifter. Above ~12 V the response changes markedly: the SW amplitude falls sharply (≈50 dB between 12 and 20 V), while the phase tends to level off and ultimately becomes ill-defined due to the very low signal-to-noise ratio. In this regime the device effectively operates as an RF switch, producing strong amplitude modulation at fixed frequency. This is partially due to the large frequency shift, which ultimately can exceed the SW bandwidth, but also to the fact that the increasing magnetoelastic field gradually breaks the single domain configuration between the two antennas. Static micro-MOKE images (Fig. 3g–i) support this interpretation: a single-domain state (Fig. 3g) persists up to ~14 V while a vertical domain wall between the antennas appears at 20 V. At 30 V the wall evolves into an S-shaped profile, consistent with increasing longitudinal magnetoelastic contributions from $\varepsilon_{xy}$. Domain formation disrupts coherent propagation and partially account for the amplitude suppression. Notably, after applying voltages that suppress transmission (~20V), reducing $V_p$ below 12 V restores the single-domain state and recoveres spin-wave transmission. These reversible, repeatable transitions indicate the device can function as an RF switch in the 12–20 V range.

Time-resolved MOKE (TR-MOKE) imaging corroborates broadband spectroscopy via real-space measurements. Experiments were performed on a nominally identical device, apart from a reduced antenna separation (20 μm) which is not relevant for MOKE experiments with optical detection. **Figure 3l** shows a $m_z$ snapshot of propagating spin waves at $V_p = 6$ V and 7 GHz under zero external field. The nearly uniform profile along the transverse direction confirms excitation of the fundamental DE mode. Linecuts from TR-MOKE maps (**Figure 3m**) reveal a systematic reduction of the wavelength with increasing $V_p$, as summarized in **Figure 3n**. This corresponds to an increase in wavevector and hence a larger accumulated phase during propagation, in excellent agreement with the phase modulation observed electrically (Fig. 3e).

**Dynamic modulation enhanced by mechanical resonance**

To probe dynamic operation, we applied a sinusoidal drive $A_p \sin(2\pi f_M t)$ to the piezo-MEMS, superimposed on a $V_{DC}$ = 10 V bias (to operate the bridge in the linear regime), and monitored the phase of $S_{21}$ at a fixed carrier frequency of 7 GHz. Under the experimentally observed linear phase response (Fig. 3e), the transmitted spin-wave phase can be expressed as $\phi(t) = \phi_0 + \beta \sin(2\pi f_M t)$, where $\beta$ is the phase modulation index proportional to the phase modulation coefficient (β = γ·A$_p$).

Experiments at low modulation frequency ($f_M = 2$ Hz) were carried out on the very same device used for VNA experiments reported in Figure 3, with an edge-to-edge distance between antennas of 30 microns. In this "slow-modulation" regime we could track the phase of $S_{21}$ in real time, as shown in Fig. 4a. The phase oscillates sinusoidally with negligible distortion and $\beta$ scales linearly with the AC amplitude A$_p$ (Fig. 4b), confirming the linear operation. A full $2\pi$ peak phase modulation ($\beta = \pi$) is obtained for A$_p$=4 V, corresponding to a phase modulation coefficient γ = ΔΦ/ΔV of 45 deg/V in nice agreement with the static coefficient derived from Fig. 3e.

The frequency dependence of $\beta$ was then measured by sweeping $f_M$ from 1 to 50 kHz, encompassing the mechanical resonance of the first bending mode. Both the electrical characterization and laser vibrometry investigation of the dynamic modes of the bridge identify a fundamental bending mode at 17.2 kHz and a higher-frequency hybrid mode near 84 kHz. However, the largest mechanical response occurs at the first bending mode, to which the magnetoelectric coupling analysis presented above directly applies. Accordingly, we concentrate on this mode in the following. Measurements were performed on a nominally identical device to that of Figure 3, but with reduced stripline antenna separation (edge-to edge 12 µm) to enhance the signal to noise ratio and enable real-time monitoring of the $S_{21}$ phase at frequencies up to 50 kHz. This was necessary because the VNA intermediate-frequency bandwidth (IFBW) could not be reduced below 1 MHz at high modulation rates, thus limiting noise filtering compared with the 1 kHz IFBW used for low-frequency (2 Hz) measurements. The modulation amplitude was $A_P=0.4$ V, while the applied DC bias was 5 V, due to some instrument constraints in these experiments. This corresponds to a piezo-MEMS operation near the lower edge of its linear regime, so the effective phase-modulation coefficient γ is expected to be reduced relative to higher bias conditions ensuring full linearity.

**Figure 4b** compares the phase modulation index $\beta(f_M)$ and the mechanical resonance curve, normalized to show their superposition. As expected also from previous works on macroscopic systems, β follows the mechanical resonance curve of the MEMS, with a clear enhancement of the modulation efficiency at resonance due to mechanical amplification of the strain induced in the CoFeB.[28,29,30] The peak phase modulation index measured at resonance in preliminary experiments is β = 35±6.5 deg for $A_P$ = 0.4 V, corresponding to γ = 87.5±16 deg/V. This is a quite high value allowing to implement a very efficient phase modulation (see discussion below), although it falls short of the ideal expectation. In a perfectly linear system the resonant enhancement of the $α_{ME}$ and γ relative to the low-frequency value should scale with the mechanical Q-factor of the resonance. Assuming as low-frequency value γ = 18 deg/V, arising from the 45 deg/V measured at 2 Hz on the device used for Fig. 3 and operated in the linear regime ($V_{DC}$=10 V) rescaled by a factor 30/12 to take into account the reduction of the distance d between the antennas, together with the experimental value Q = 31, the expected γ should be 540 deg/V. The device's deviation from perfect linearity near $V_{DC}$ = 5 V, together with mechanical coupling to a nearby second bridge (Fig. 1a) forming a pair of coupled oscillators, may account for this discrepancy. However, we point out that the corresponding converse magnetoelectric coefficient measured at resonance is on the order of 6 mT/V, higher than maximum values reported in laminated composites.[28]

**Discussion**

Our results demonstrate that magnetoelastic coupling in a strain-engineered MEMS platform enables efficient electrical control of spin-wave propagation for RF signal processing. The device operates either with an external magnetic field - extending the available frequency tuning range - or in a standalone mode that exploits growth-induced transverse anisotropy to set a remanent single-domain configuration. Reliable operation was demonstrated in the standalone mode for spurious external fields < 1.5 mT, but further robustness can be achieved by increasing uniaxial anisotropy or introducing auxiliary biasing mechanisms such as exchange bias. Noteworthy, this mode allows compact integration (the device footprint is 2000x600 μm and could be further reduced in optimized designs) and no need of external biasing magnets. Furthermore, both voltage-controlled RF switching (≈50 dB attenuation) and high-efficiency phase-shifter operation (large phase-modulation coefficients) can be implemented.

We demonstrate highly efficient phase modulation in both static and dynamic regimes. In static operation we measured a phase-modulation coefficient up to ≈ 60 deg/V, with ±10% variability depending on the piezo pooling procedure. Dynamically, the modulation efficiency increases markedly near the fundamental bending resonance (17 kHz), with the phase-modulation coefficient tracking the mechanical response and reaching γ = 87.5 ± 16 deg/V at resonance. Although below the ideal theoretical maximum, this γ enables a full 2π phase swing of a 7 GHz carrier with only ≈2.2 V drive, well within CMOS voltage levels. Operating the device fully in the linear regime and reducing the coupling to adjacent MEMS would allow to reach the theoretical amplification γ by the mechanical Q-factor. Extrapolating from our low-frequency γ (measured at 2 Hz, Fig. 4a), a 2π phase modulation at resonance could be achieved with a drive amplitude as small as 130 mV.

Beyond the fundamental bending mode, additional mechanical resonances provide further degrees of freedom. Figures 4c–e illustrates the case of the second torsional mode at ~84 kHz. In the present symmetric actuation scheme, this mode is only weakly excited through hybridization, since it is antisymmetric with respect to $y$ while the piezo actuation is symmetric. However, independent addressing of the four piezo pads would enable an efficient excitation. In this case, the dominant strain component is $\varepsilon_{xy}$, producing a longitudinal magnetoelastic field component $H_{ME,x}$. For a transverse equilibrium magnetization this induces a small tilt of the magnetization toward $x$, but even modest canting suffices to yeld sizable phase modulation of propagating waves. Note that $H_{ME}$ flips sign with the sign of $\varepsilon_{xy}$, but the resulting change in band dispersion is insensitive to the tilt direction (left or right) relative to the wavevector; hence the phase modulation appears at twice the mechanical drive frequency (≈168 kHz). This example highlights how mode symmetry and targeted strain engineering can be used to tailor the spatial profile and temporal response of the magnetoelastic field.

A key advantage of our approach over conventional tuning based on magnetic fields is low power consumption. In static operation, once the bridge is displaced, power is limited to piezo leakage currents (<1 nA), corresponding to intrinsic nanowatt dissipation. Under resonant dynamic actuation (400 mV at 17.2 kHz), the dissipated power can be calculated assuming the typical tanδ of PZT (on the order of $10^{-2}$) and using the formula for a dielectric with losses.[36,37] For the measured device capacitance of 2.4 nF, this leads to an intrinsic power dissipation of about 200 nW that would become ≈ 6 µW in case of a 2.2 V drive amplitude ensuring 2π modulation. Achieving a 2π phase modulation by uniform magnetic-field modulation would require an AC field amplitude of about 6 mT (see Figure 3e). The corresponding power dissipation of an electromagnet or an integrated coil providing that field would be orders of magnitude higher. In the assumption of placing a 400 µm long, 60 µm wide, 1 µm thick copper current line beneath the magnonic conduit, producing a magnetic field of 6 mT would require 600 mA of current and local dissipation of ≈40 mW by thermal dissipation alone. Finally note that the proposed architecture is best suited to implement tunable and relatively "slow" analog RF functions, such as phase shifting, adaptive filtering, switching and phase modulation, crucial, for instance, in the implementation of phase-arrays for RF antennas and Frequency-Modulated Continuous-Wave Radars.[38,39] Its application in high-speed digital modulation, requiring simultaneous amplitude and phase control at nanosecond timescales, is prevented by the "slow" mechanical response of piezo-MEMS (switching times of tens–hundreds of microseconds). Nevertheless, conventional RF MEMS are widely used in microwave electronics alongside monolithic microwave integrated circuits (e.g., GaAs and CMOS switches), particularly for applications demanding low loss and continuous (non-digital) RF modulation.[40,41] Although this paper presents a proof-of-concept device that is not yet optimized across all RF-component metrics, it demonstrates the practical potential of exploiting the broad, easily tunable response of spin waves in realistic RF components.

**Conclusions**

In this paper we demonstrated an efficient, integrated approach for the modulation of RF signals carried by spin waves through magnetoelastic coupling in a strain-engineered MEMS platform. The usage of a CoFeB waveguide integrated on a suspended piezoelectric bridge enables large magnetoelectric coefficient of ~1 mT/V. At 7 GHz, this produces a phase shift up to ~730 deg for 14 V applied, with a phase modulation coefficient of 62 deg/V in the linear regime. For higher voltages, transmission is suppressed by ~50 dB due to both the large shift of the frequency band and the micro magnetic reconfiguration, enabling the same device to operate also as a voltage-controlled RF switch. Exploiting the mechanical resonance of the MEMS we demonstrate the possibility of reaching a 2π phase modulation of a 7 GHz carrier signal with just 2.2 V drive amplitude and power-consumption of a few µW. Higher efficiency could be obtained in optimized devices, fully exploiting the maximum theoretical amplification of the phase modulation efficiency by the Q-factor of the mechanical resonance. Our monolithic piezo-magnonic architecture combines large phase/amplitude tunability and low dynamic power consumption in a compact platform, highlighting the potential of hybrid piezo- magnonics for reconfigurable, low-power RF signal processing.

**Methods**

***Device fabrication*** Device fabrication was performed on the piezoelectric MEMS bridges kindly provided by STMicroelectronics. The magnonic waveguide was first patterned on the suspended bridge by optical lithography, followed by magnetron sputtering under an in-plane magnetic field of 27 mT and lift-off, using a Ta(5 nm)/CoFeB(40 nm)/Ta(5 nm) trilayer stack. Subsequently, a 500-nm-thick $SiO_2$ layer was deposited by sputtering to electrically isolate the waveguide from the RF antennas used for Oersted-field excitation and to passivate the exposed metallic contacts of the piezoelectric MEMS. During this step, the MEMS contact pads were kept exposed using a lift-off-defined mask. Finally, RF excitation antennas were fabricated by optical lithography, electron-beam evaporation, and lift-off of a Ti(10 nm)/Au(100 nm) bilayer.

***Simulations and modelling*** Mechanical and electrical simulations of the MEMS were performed using COMSOL Multiphysics 6.3 to determine the device deformation under the application of a voltage and the resulting strain distribution. These simulations were used to identify the spatial regions of maximum strain and thus guide the placement of the magnetostrictive element. COMSOL was also employed to compute the eigenfrequencies of the structure and to identify its mechanical resonance modes. Micromagnetic simulations were carried out using Ubermag with the OOMMF solver to determine the equilibrium magnetization state of the magnonic conduit under the strain field simulated by COMSOL.

***MEMS characterization*** The MEMS electro-mechanical response was acquired using a HF2LI Lock-in amplifier together with its transimpedance amplifier (HF2TIA). The lock-in was used to apply a voltage to either the top or bottom electrode of the device while the other electrode was connected to the virtual ground node of the HF2TIA. The frequency of the applied voltage was swept to reconstruct the electromechanical response of the device at resonance and estimate its quality factor. For each frequency, the current produced by the capacitance and the displacement of the piezoelectric actuators was converted into a voltage by the HF2TIA and sent to the lock-in amplifier that reconstructed the frequency response of the device impedance. After subtracting the capacitive contribution to the detected signal, the data was fitted using the analytical response of a damped harmonic oscillator. The key features of the mechanical resonator derived from electric measurements were further confirmed using a Digital Holographic Microscope from Lyncetec.

***Broadband spectroscopy*** Broadband spin-wave spectroscopy was performed using a RF probe station equipped with a quadrupolar vector electromagnet providing in-plane magnetic fields up to 200 mT, and a four-port vector network analyzer with a bandwidth up to 43 GHz (R&S ZNA43). Measurements were carried out using either frequency-swept or time-sweep acquisition modes. In frequency-swept measurements, the excitation frequency was swept from 4 MHz to 40 GHz using 10,000 points, resulting in a harmonic frequency grid [δf,2δf,3δf,…,Nδf] with δf=4 MHz. This uniform spacing enables time-gating procedures. The applied RF power was 0 dBm, and the full scattering matrix was recorded. For experiments investigating strain-induced effects (Fig. 2a–f), a predefined measurement protocol was adopted to minimize hysteresis-related artifacts. The MEMS piezoelectric actuator was first polarized by applying a high voltage, after which the voltage was reduced to the lowest value of the investigated range. The voltage was supplied using a Keithley 2450 source meter. For each voltage value, measurements were performed by sweeping the magnetic field from the lowest to the highest value and acquiring the scattering matrix over the harmonic frequency grid. Only after completing the full magnetic-field sweep was the voltage increased to the next value. This procedure ensures that all measurements performed at a given voltage correspond to an identical mechanical state of the MEMS, thereby suppressing effects arising from mechanical hysteresis. Measurements targeting the dynamic modulation of spin-wave propagation were conducted using a modified electrical configuration for the MEMS actuation. In this case, the MEMS bottom electrode was connected either to ground or, during resonance measurements (such as the black trace in Fig. 4c), to the input of the lock-in amplifier as described in the previous paragraph, while the top electrode was connected to the output of a voltage amplifier driven by the lock-in amplifier. This configuration enables the application of a DC bias voltage with a superimposed AC modulation at a selected frequency. The spin-wave response was recorded in time-sweep mode, i.e. by measuring the scattering matrix at a fixed frequency as a function of time. Provided that the sampling rate fulfils the Nyquist criterion for the applied modulation frequency, this approach allows direct tracking of the time-dependent modulation of the spin-wave signal induced by the dynamically varying strain.

***Static and dynamic MOKE measurements*** Static magneto-optical Kerr effect (MOKE) measurements were performed using a home-built microMOKE setup using an optical microscope operated in the longitudinal configuration. Illumination was provided by white p-polarized light, and the reflected beam was analyzed using an s-oriented polarizer to isolate the Kerr rotation signal before detection with a high-sensitivity, high-resolution charge-coupled device camera. The images were acquired using a 50x objective, allowing the capture of areas of approximately 150 by 400 micrometers with a spatial resolution of about 500 nm. Stroboscopic images of the magnetization dynamics upon excitation of SWs have been recorded using a Tr-MOKE (Time-Resolved Magneto-Optical Kerr Effect) apparatus exploiting a train of polarized laser pulses (638 nm, pulse-width of 70 ps) which strike the magnetized surface and undergo a polarization rotation, proportional to the oscillating out-of-plane (OOP) magnetization component. The frequency of the spin wave and the repetition rate of the laser pulse train are linked by an integer ratio, so that each laser pulse probes the wavefront always at the same phase of the oscillation. In this way, the system works like a stroboscope, "freezing" the ultrafast motion of the wave at a precise instant in its cycle. 2D maps are obtained by scanning the sample with a piezo-actuator under the laser beam, which is focused into a spot with a diameter smaller than 1 micron.

**Acknowledgements**

The authors acknowledge funds from the European Union via the Horizon Europe project "MandMEMS", grant 101070536 and from NextGenerationEU, PNRR MUR – M4C2 – Investimento 3.1, project IR_0000015 – "Nano Foundries and Fine Analysis – Digital Infrastructure (NFFA–DI)", CUP B53C22004310006. We thank: D. Ferling, from Nokia-Nell-Labs for fruitful discussions on the potential of our devices as RF components; P. Pirro for general discussions on the applications of our platform; M. Vitali for technical support in MOKE experiments; L. Castoldi and S. Zerbini from STMicroelectronics for providing MEMS devices. In the framework of the JRP STEAM. This work has been partially performed at Polifab, the micro and nanofabrication facility of Politecnico di Milano.

**Author contribution**

Maria Cocconcelli initially proposed the concept of the hybrid device. M.C and A. A. mainly contributed to device fabrication, characterization and data analysis. N. P. performed the static micro-MOKE characterization. P.F. carried out the TR-MOKE analysis. F.M. performed the characterization of the MEMS mechanical resonance. R.B wrote the manuscript with the collaboration of all the authors. F.M and R.B supervised the research.

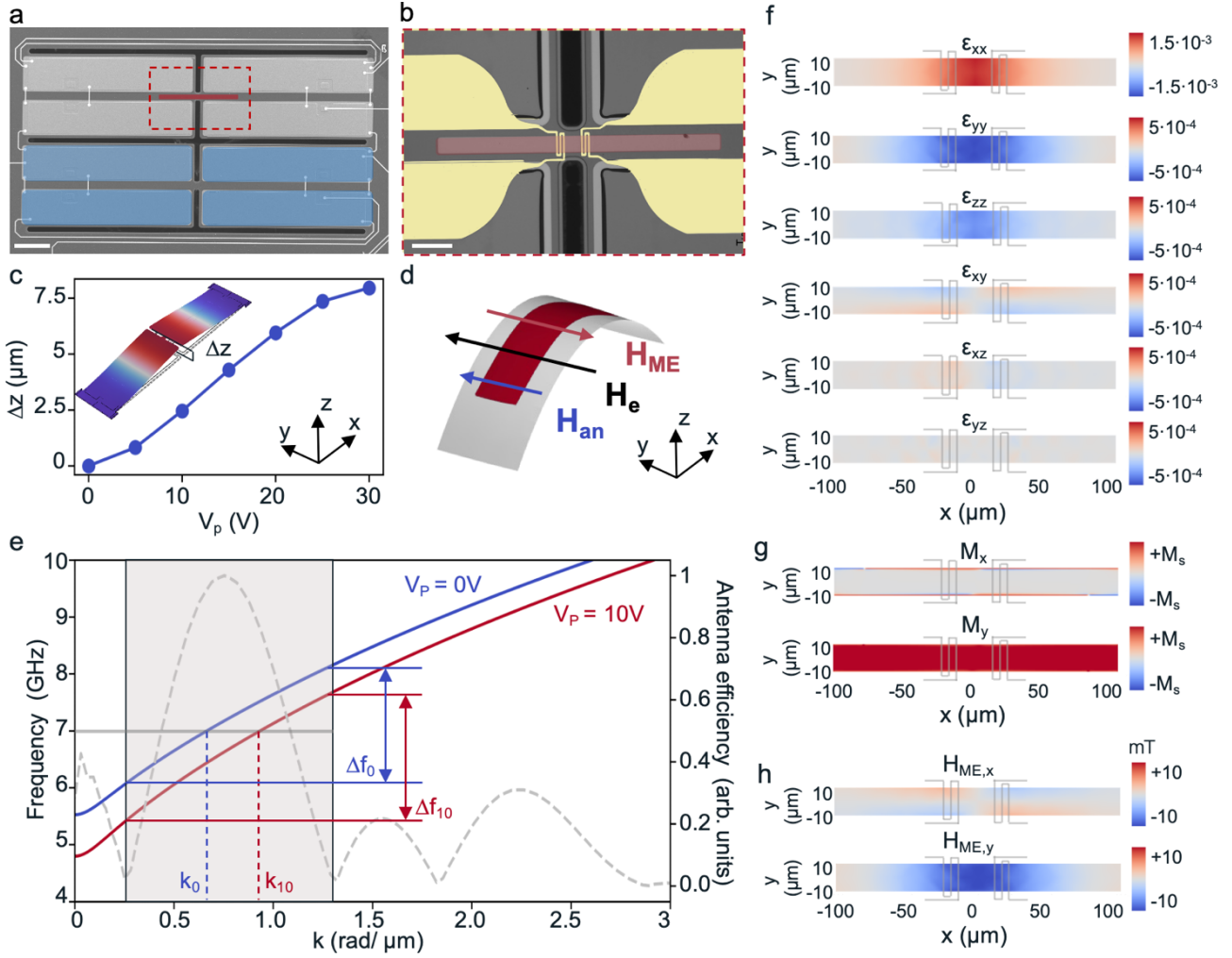

**Figure 1:** a) Top view of two piezo-bridges (scale-bar 200 µm). b) zoom of the central part of a single bridge (scale-bar 50 µm) with the CoFeB waveguide (red) and meander antennas. c) Vertical displacement of the bridge vs. voltage $V_p$ applied to the piezo. d) Sketch of the bended CoFeB conduit with consistent orientations of the fields. e) DE band dispersions of SW for the unstrained ($V_p$=0V – blue continuous line) and strained ($V_p$=10 V - red continuous line) condition, together with the antenna excitation efficiency (dashed line). $\Delta f_0$ and $\Delta f_{10}$ represent the bandwidth of SW corresponding to the first peak of the excitation efficiency of the antenna for 0 and 10 V. $k_0$ and $k_{10}$ are the corresponding wave-vectors excited at 7 GHz. f) Color map of the strain field generated by $V_p$=10 V. g) Corresponding configuration of the magnetization in the CoFeB conduit for zero external field (remanence). h) Color map of the calculated magnetoelastic field $H_{ME}$ for $V_p$=10 V.

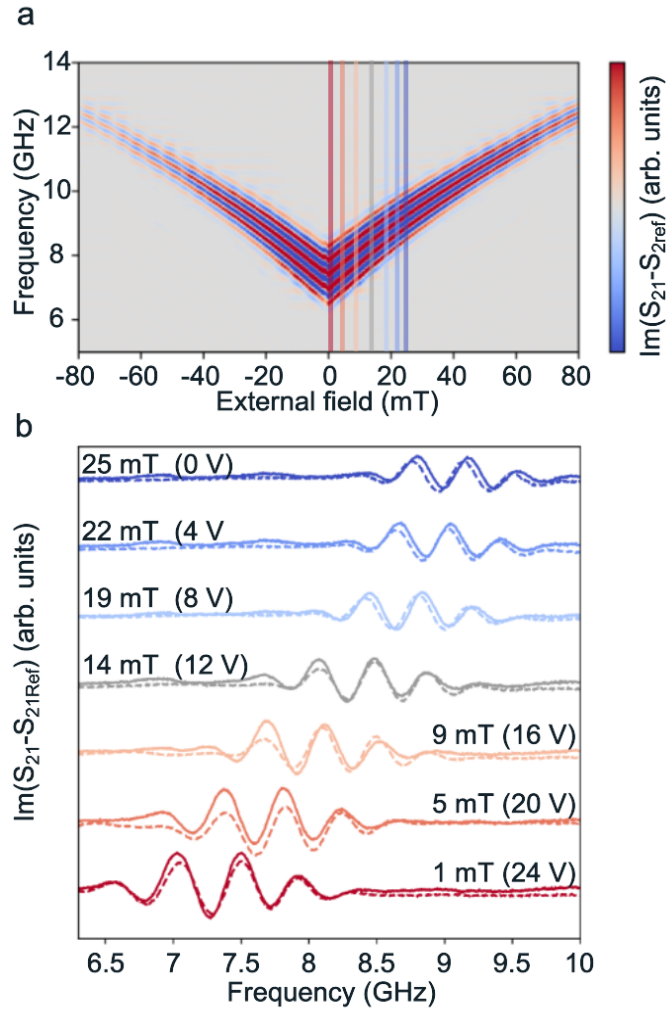

**Figure 2:** a) Colour map of the reference subtracted imaginary part of the scattering coefficient S21 from VNA experiments, as a function of frequency and external magnetic field, for $V_p$=0 V. b) Continuous lines: linecuts of the plot in Fig. 2a, corresponding to the vertical lines in panel a, for specific values of the external field $H_e$. Dashed lines: spectra taken at fixed $H_e$=25 mT and variable $V_p$ (values within brackets) allowing to reproduce the continuous lines, i.e. to obtain the same effective field because of the creation of a negative magnetoelastic field.

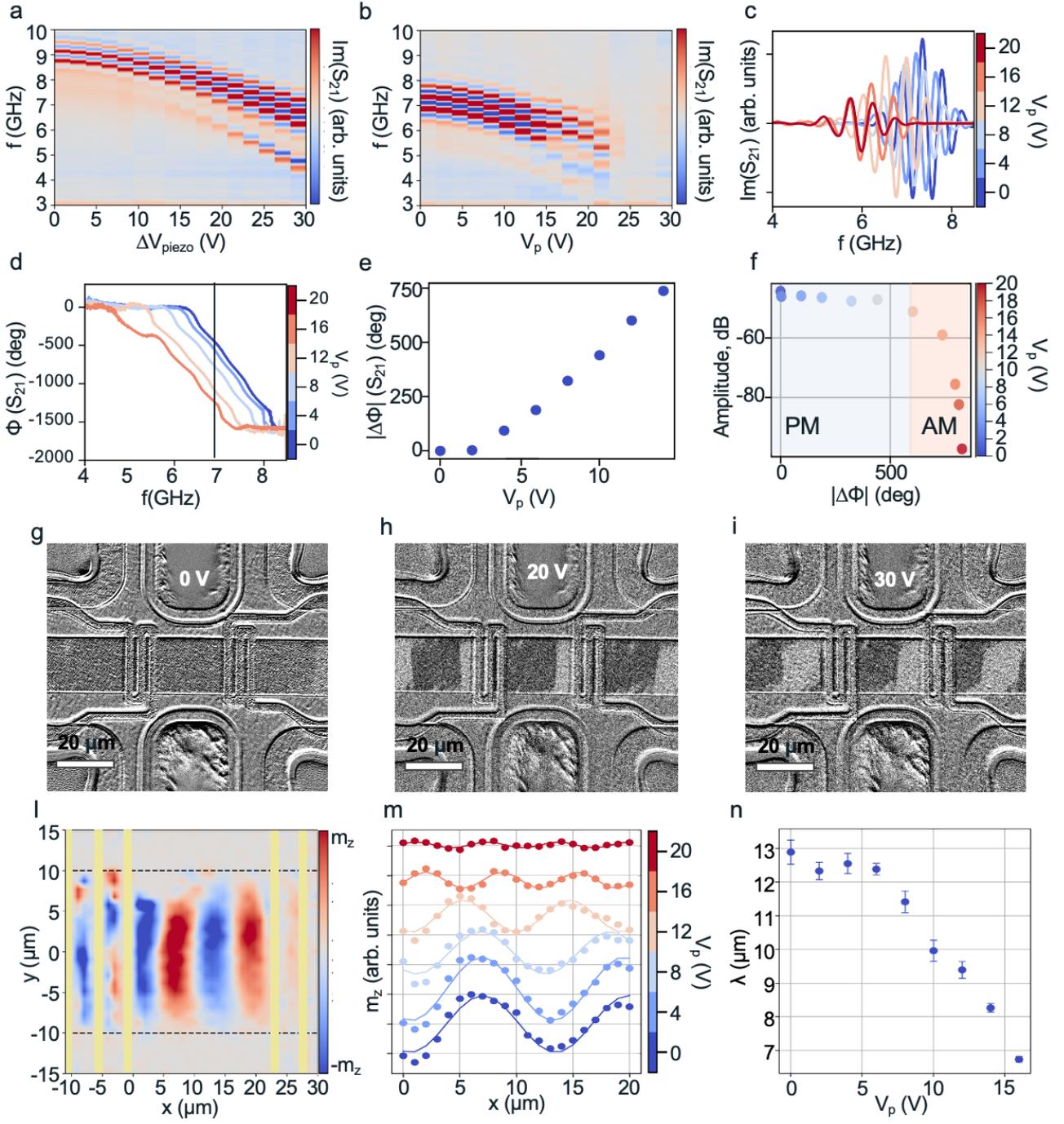

**Figure 3**: a) Colour plot of the reference subtracted imaginary part of $S_{21}$ vs. frequency and applied voltage at fixed $H_e$=25 mT. b) Same as in a, but for zero external magnetic field. c) Linecuts of panel 3b for selected values of $V_p$ between 0 and 20 V in steps of 4 V. d) Phase of $S_{21}$ vs frequency for $V_p$ in the 0-16 V range where the phase at 7 GHz is well defined. e) Relative phase shift induced by $V_p$ at 7 GHz. f) Operational diagram of the device, showing for each value of $V_p$ between 0 and 20 V the corresponding amplitude and phase modulation. The light-blue and light-pink areas correspond, respectively, to the "phase shifter" and "RF switch" operation modes. g,h,i) Static MOKE images of the region of the CoFeB waveguide with antennas at zero $H_e$ and $V_p$ = 0, 20, 30 V. l) TRMOKE image of the dynamic part of the local magnetization $m_z$ showing SW wavefronts generated by the left antenna at 7 GHz for $V_p$= 6 V. m) Linecuts of $m_z$ maps along the center of the waveguide for different values of $V_p$. n) Plot of the experimental values of the SW wavelength ($\lambda$) vs. Vp extracted from data in panel m.

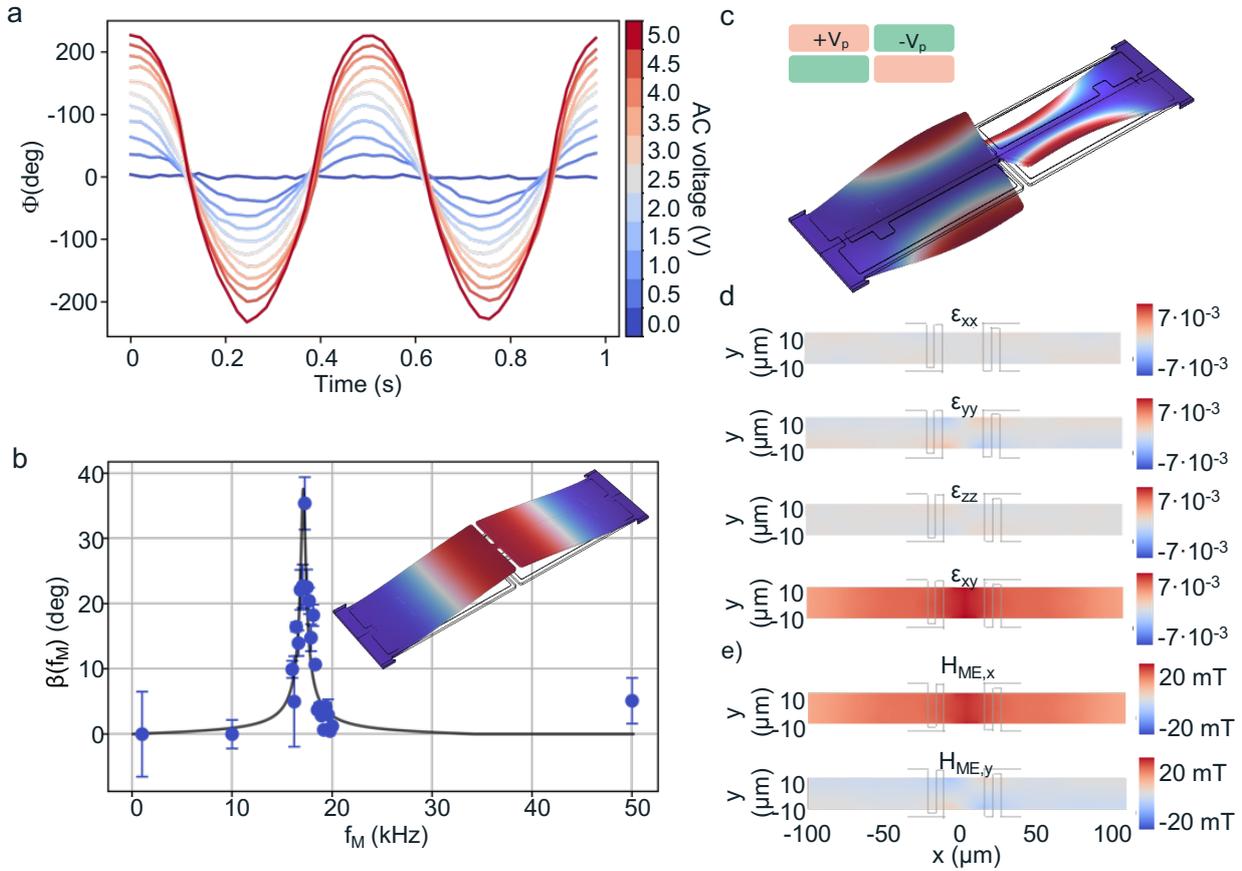

**Figure 4**: a) Real-time evolution of the phase of $S_{21}$ at 7GHz, for $H_e$=0mT and different amplitudes of an AC signal at 2 Hz applied to the piezo. b) Blue dots: phase modulation coefficient β vs. the modulation frequency of the AC signal applied to the piezo. Continuous line: mechanical resonance curve. In the insert the spatial profile of the first bending mode corresponding to the observed resonance. c) Spatial profile of the second torsional mode that could be efficiently excited at about 84 kHz for an asymmetric application of voltages to the four piezoelectric elements, as shown in the inset. d) Color maps of the strain field produced by this mode, for Vp=10V. e) Corresponding maps of the magnetoelectric field produced.